\documentclass[journal=jacsat,manuscript=article]{achemso}

\usepackage[version=4]{mhchem} 
\usepackage{textcomp}
\usepackage{upgreek}

\author{Yixiong Ji}
\affiliation{ARC Centre of Excellence in Exciton Science, School of Chemistry, University of Melbourne, Victoria 3010, Australia}

\author{Paul Mulvaney}
\email{mulvaney@unimelb.edu.au}
\affiliation{ARC Centre of Excellence in Exciton Science, School of Chemistry, University of Melbourne, Victoria 3010, Australia}

\title[An \textsf{achemso} demo]
{A "Redox-free" Synthesis of CZTS Nano Ink}
\abbreviations{Solar energy materials}
\begin{document}

\begin{tocentry}

Some journals require a graphical entry for the Table of Contents.
This should be laid out ``print ready'' so that the sizing of the
text is correct.

Inside the \texttt{tocentry} environment, the font used is Helvetica
8\,pt, as required by \emph{Journal of the American Chemical
Society}.

The surrounding frame is 9\,cm by 3.5\,cm, which is the maximum
permitted for  \emph{Journal of the American Chemical Society}
graphical table of content entries. The box will not resize if the
content is too big: instead it will overflow the edge of the box.

This box and the associated title will always be printed on a
separate page at the end of the document.

\end{tocentry}


\newpage
\begin{abstract}

\hspace{1em}A large open-circuit (V$_{oc}$) deficit restricts current kesterite device performance. The primary challenge is to achieve control over the phase composition and purity of the kesterite absorber. This is hampered by the fact that the metals copper and tin have multiple valence states and this leads inevitably to the formation of multiple phases. Specifically for solution-based fabrication procedures for kesterite, the pursuit of phase purity extends to the synthesis of CZTS precursor solution or nanoparticle dispersed inks (nano inks). In this work, a "redox-free" synthesis of CZTS nano ink is developed by mixing metal precursors with careful valence state control in non-toxic solvents. The issue of secondary phase formation during the synthesis process of kesterite is effectively resolved. Additionally, molecular solutions and nanoparticle inks with identical compositions exhibit significantly different abilities in phase control. Nanoparticles pre-synthesized in the solution state exhibit superior phase control by following a more ideal phase formation path. This provides a new pathway for the synthesis of kesterite with unprecedented control of the phase composition and purity.

\end{abstract}

\newpage
\section{Introduction}
 
\hspace{1em}Utilizing solar energy with efficient strategies is one of the crucial scientific challenges of this century. During the past decades, exciting progress in photovoltaic technologies has been achieved, but sustainable, low-cost, and toxic-free absorber materials are still in demand. Benefiting from its earth-abundant compositions and excellent optoelectronic properties, kesterite Cu$_{2}$ZnSn(S, Se)$_{4}$ (CZTSSe) is emerging as a competitive candidate in inorganic absorber materials for solar cells. Similar to other members of the metal chalcogenide family like CdTe and its counterpart Cu(In, Ga)Se$_{2}$ (CIGS) absorbers, the property of the CZTSSe absorber together with the photovoltaic performance of the solar cells highly depends on the phase-purity of the as-deposited absorber layer \cite{Gong2022elemental,Zhou2023control}. 

Currently, the enhancement of the kesterite device performance relies on the rapid improvement of open-circuit voltage. The primary causes of V$_{oc}$ loss are the low phase purity of the absorber material and the presence of deep-level defects \cite{Secondaryphases, Kumar2015strategic}. Existing solutions predominantly focus on the precise adjustment of CZTS components, including cation and alkaline metal doping, promoting the efficiency improvement from 12.6\% to over 13\% \cite{Agdoping2016optoelectronic, SuCd2020device, Yang2024multiple, HaoXin202312}. Consequently, solution-based preparation methods, which facilitate better control over composition and phase formation, offer significant advantages over vacuum-based techniques. Molecular solutions incorporate all of the constituents at a molecular level leading to the in-situ formation of desired compositions and phases after heat treatment. Various strategies like hydrazine-based \cite{Wang2014device}, DMSO-based \cite{HaoXin202312}, amine-thiol-based \cite{Fujunjie2018tuning}, and 2-methoxyethanol-based \cite{Zhao2021ambient} solution processes have been successfully employed for fabricating CZTSSe absorber layers. 

The pursuit of controlling phase purity has led researchers to focus on the timing of the appearance of the kesterite phase and the transformation pathways from the precursor to the kesterite phase. Consequently, research has gradually shifted to the critical phase transition stage: the high-temperature selenization process \cite{Zhou2023control, Ji2024single}. However, being a black-box reaction process presents significant challenges in monitoring and control. Therefore, we propose returning to phase control research in the solution state. By making use of wet chemical methods under low-temperature, solution-state conditions, which are easier to monitor and control, we aim to achieve precise phase control of the final product by pre-synthesizing the target phase nanoparticle in the precursor solution. The synthesized CZTS nanoparticle inks can be directly used to prepare CZTS precursor films, which can then be converted into CZTSSe absorber layers through a conventional selenization process.

Particle-based approaches typically require long-chain ligands and organic agents to ensure wetting and particle dispersion and to avoid film cracks and delamination. Time-consuming purification processes are also inevitable. The aforementioned challenges are also the reason why nanoparticle-based solutions, the first-ever solution-based method for depositing CZTS, are gradually being replaced by more easily handled molecular solution methods in the preparation of CZTSSe solar cells \cite{Guo2009NP}.

Herein, we propose a low-toxic CZTS nano-ink through a "redox-free" synthesis route by mixing metal complexes with target valence in one pot. Typically, water-based thioglycolic acid (TGA) and n-butylamine (BA) work as cosolvents and as a pH buffer, while Cu(I)/Zn oxides can form hydrophilic metal-S complexes when added to this solution. A molecular-level Sn source from tin metal chalcogenide complex (Sn-MCC) was synthesized by dissolving tin and sulfur powder into aqueous ammonium sulfide, further stabilized by protonated BA (H$_{3}$N$^{+}$CH2CH2CH3) after mixing. All the metal cations are self-stabilized in an as-synthesized CZTS precursor solution with perfect valence by complexing. In addition to this, Sn-MCC works as a ligand to form the Cu-Zn-Sn-S nanoparticles through a temperature-dependent ligand exchange process. Deposition using this nano-ink successfully avoids the formation of secondary phases by engineering the valence states of the metal precursors, nd yields pure kesterite CZTSSe absorbers after selenization. Lastly, compared to a molecular solution with the same composition, the pre-synthesized nanoparticle solution has offers lower energy cost and avoids the need for an energy-consuming phase transition from wurtzite to kesterite.

\newpage
\section{Results and discussion}

\subsection{Design Principles for Self-stabilized CZTS Hybrid Nano-inks}

\begin{figure}[ht]
    \centering
    \includegraphics[width=\textwidth]{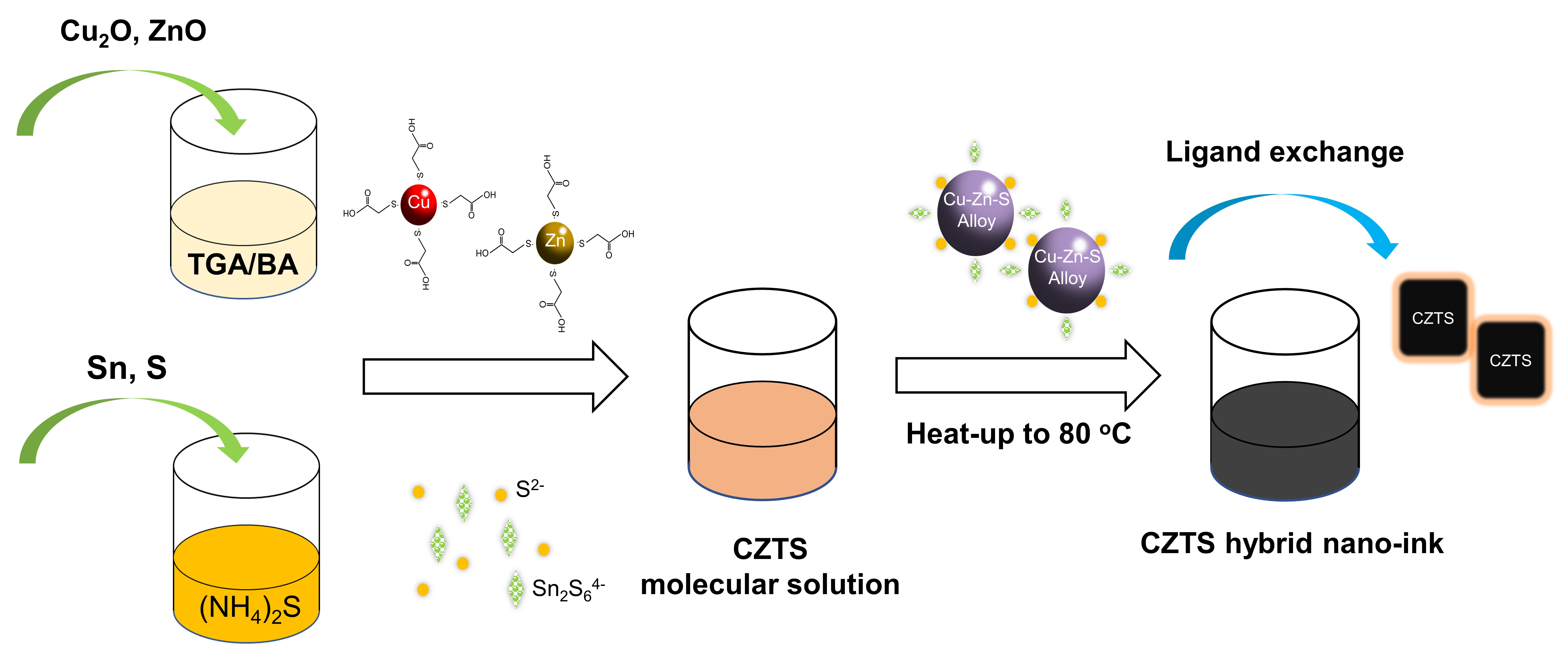}
    \caption[Schematic process of the synthesis of hybrid CZTS nano-ink.]{Schematic process of the synthesis of hybrid CZTS nano-ink.}
    \label{Synthesis}
\end{figure}

\hspace{1em}CZTS nano-inks were synthesized via a redox-free pathway by mixing three metal complexes in one pot. The process is shown in Figure \ref{Synthesis}. Cuprous oxide(I) and Zinc oxide(II) were added into the mixture of thioglycolic acid (TGA) and butylamine (BA) as metal sources, forming M(SCH$_{2}$COOH)$_{x}$ complexes \cite{Tian2014versatile}. After mixing with the Sn-MCC solution, a slightly orange CZTS molecular solution was made. Heating this molecular solution led to the formation of yellow Cu(I)-Zn-S alloy nanoparticles, which were self-stabilized by Sn-MCC. When heating to 80 $^{O}$C, thiol-amine groups on the metal ions were partially substituted by Sn-MCC, which is widely used as an inorganic ligand with a high coordinating affinity in nanoparticle synthesis, forming black CZTS nanoparticles in solution \cite{Jiang2012SnMCC, Zhong2014one}. 

A STEM image of the collected CZTS nanoparticles is shown in Figure \ref{STEM} and reveals an average diameter of around 50 nm. The absorption spectrum and calculated optical bandgaps are shown in Figure \ref{abs photos}, confirming that the black CZTS nano-inks have a band gap of around 1.58 eV. We observed that the ligand exchange process does not occur at room temperature, meaning a stable CZTS molecular solution can be made and stored with the same composition. The CZTS nano-ink can readily be made on multigram scales. The elemental ratio of Cu/(Zn + Sn) = 0.8 and Zn/Sn = 1.24 were used for our copper-deficient nano-ink synthesis. No further purification process is needed. The as-fabricated ink was found to be stable for months.

\begin{figure}[H]
    \centering
    \includegraphics[width=1.05\textwidth]{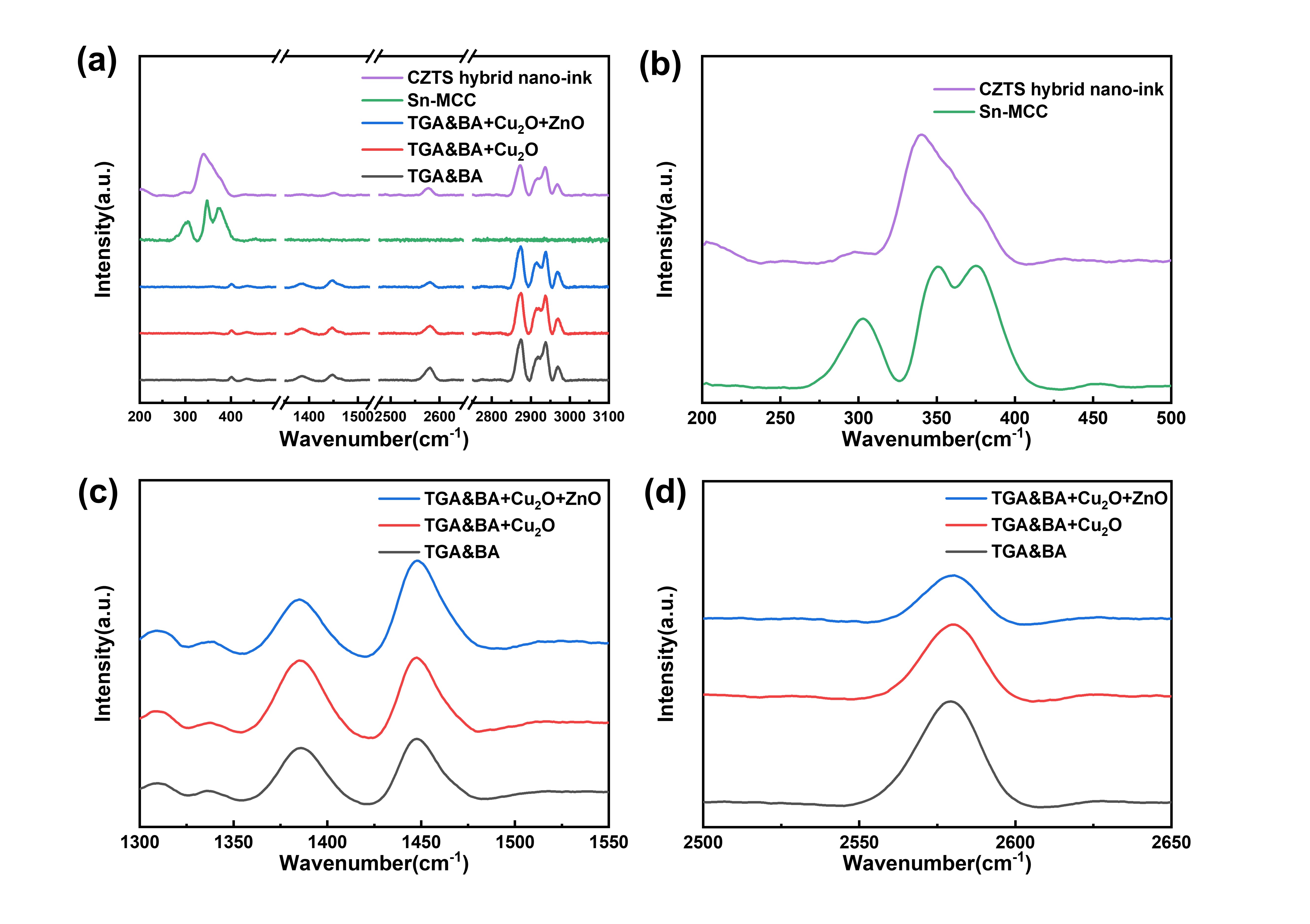}
    \caption[Raman spectra of liquid CZTS nanoparticle samples.]{Raman spectra of coordination structure variations. (a) Raman spectra of precursor solutions and the synthesized CZTS hybrid nano ink. Signals corresponding to COO$^{-}$ stretching and SH groups are amplified and displayed in (c) and (d) respectively. (b) Amplified Raman signals of Sn-MCC and CZTS are shown in (b). All signals are normalized by the intensity of CH$_{2}$ peaks from amines.}
    \label{Liquid Raman}
\end{figure}

The chemical reactions involved in CZTS formation were investigated by liquid Raman spectroscopy, as shown in Figure \ref{Liquid Raman}. The full range of Raman absorption bands have also been shown here, and separated into four parts for parsing. In the spectra of the TGA/BA solution, characteristic absorption bands for both the released SH and COO$-$ groups are observed at $~$2570 cm$^{-1}$ and $~$1448 cm$^{-1}$, respectively. \cite{Raman1448, Raman2570} Changes in the absorption bands are observed after adding the metal oxides. On one hand, a significant decline in the SH absorption band is evident in (d) when Cu$_{2}$O and ZnO were added into the mixture consecutively, which implies that the SH group in TGA was deprotonated by BA, leaving S atoms to coordinate with copper or zinc to form metal complexes \cite{Guo2020coordination}. On the other hand, it is indicated that the COOH group was also deprotonated with the absence of the peaks from COO${-}$, but showed less prone coordination with metal atoms due to the seldom changed intensity in (c). Notably, although the competition relationship between metal${-}$S and metal${-}$O in TGA when coordinating with metal atoms, the carboxyl group in TGA plays a minor role in dissolving metal oxides, which is also demonstrated by the poor solubility of ethyl thioglycolate and ethanedithiol \cite{Tian2014versatile}. Nonetheless, carboxyl in TGA together with amine groups in BA, working as hydrophilic groups, promises a good solubility of metal oxides in Ethanol-water-based metal precursors \cite{Qumechanism}. A possible dissolution mechanism is shown in equations \ref{equation 3-1},\ref{equation 3-2}, \ref{equation 3-3}.

\newpage
\begin{equation}
\begin{split}
\mathrm{HSCH}_{2} \mathrm{COOH}+\mathrm{CH}_{3} \mathrm{CH}_{2} \mathrm{CH}_{2} \mathrm{NH}_{2} &\rightarrow{^{-}}\mathrm{SCH}_{2} \mathrm{COOH}+\mathrm{HSCH}_{2} \mathrm{COO}^{-} \\ &+\mathrm{H}_{3} \mathrm{~N}^{+} \mathrm{CH}_{2} \mathrm{CH}_{2} \mathrm{CH}_{3} 
\end{split}
\label{equation 3-1}
\end{equation}
\begin{equation}
\begin{split}
\mathrm{Cu}_{2} \mathrm{O}+\mathrm{HSCH}_{2} \mathrm{COOH}+\mathrm{CH}_{3} \mathrm{CH}_{2} \mathrm{CH}_{2} \mathrm{NH}_{2} &\rightarrow \mathrm{Cu}\left(\mathrm{SCH}_{2} \mathrm{COOH}\right)_{\mathrm{x}}\\&+\mathrm{H}_{3} \mathrm{~N}^{+} \mathrm{CH}_{2} \mathrm{CH}_{2} \mathrm{CH}_{3}+\mathrm{H}_{2} \mathrm{O}
\end{split}
\label{equation 3-2}
\end{equation}
\begin{equation}
\begin{split}
\mathrm{ZnO}+\mathrm{HSCH}_{2} \mathrm{COOH}+\mathrm{CH}_{3} \mathrm{CH}_{2} \mathrm{CH}_{2} \mathrm{NH}_{2} &\rightarrow \mathrm{Zn}\left(\mathrm{SCH}_{2} \mathrm{COOH}\right)_{\mathrm{x}}\\&+\mathrm{H}_{3} \mathrm{~N}^{+} \mathrm{CH}_{2} \mathrm{CH}_{2} \mathrm{CH}_{3}+\mathrm{H}_{2} \mathrm{O}
\end{split}
\label{equation 3-3}
\end{equation}

\begin{equation}
\mathrm{Cu}^{+}+\mathrm{Sn}^{4+} \leftrightarrow \mathrm{Cu}^{2+}+\mathrm{Sn}^{2+}
\label{equation 3-4}
\end{equation}
\begin{equation}
\begin{split}
\mathrm{Cu}^{2+}+\mathrm{HSCH}_{2} \mathrm{COOH} &\rightarrow\left(\mathrm{HSCH}_{2} \mathrm{COO}\right) \mathrm{Cu}(\mathrm{I})\\&+\mathrm{HOOCCH}_{2} \mathrm{~S} \cdot \mathrm{SCH}_{2} \mathrm{COOH}+\mathrm{H}_{2} \mathrm{O} 
\end{split}
\label{equation 3-5}
\end{equation}
\begin{equation}
\begin{split}
\mathrm{Sn}^{2+}+\mathrm{HOOCCH}_{2} \mathrm{~S} \cdot \mathrm{SCH}_{2} \mathrm{COOH} \rightarrow \mathrm{Sn}\left(\mathrm{SCH} \mathrm{SOOO}_{2}(\mathrm{IV})+\mathrm{H}_{2} \mathrm{O}\right.
\end{split}
\label{equation 3-6}
\end{equation}
\begin{equation}
\begin{split}
\mathrm{Sn}+\mathrm{S}+\left(\mathrm{NH}_{4}\right)_{2} \mathrm{~S}\rightarrow \mathrm{NH}_{4}^{+}+\left[\mathrm{Sn}_{2} \mathrm{~S}_{6}\right]^{4-}
\end{split}
\label{equation 3-7}
\end{equation}
\begin{equation}
\begin{split}
{\left[\mathrm{Sn}_{2} \mathrm{~S}_{6}\right]^{4-}+\mathrm{H}_{3} \mathrm{~N}^{+} \mathrm{CH}_{2} \mathrm{CH}_{2} \mathrm{CH}_{3} \rightarrow\left[\mathrm{Sn}_{2} \mathrm{~S}_{6}\right]\left(\mathrm{H}_{3} \mathrm{NCH}_{2} \mathrm{CH}_{2} \mathrm{CH}_{3}\right)_{4}}
\end{split}
\label{equation 3-8}
\end{equation}

Considering further the material characteristics of the multi-valence nature of Cu and Tin elements, reversible redox reactions as described in equation \ref{equation 3-4} are common and even exist in state-of-the-art precursor solutions\cite{Gong2021Sn2+4+}. Here, these reactions are subtly suppressed from left to right by equation \ref{equation 3-5} and equation \ref{equation 3-6}. Unwanted Cu(II) can be reduced to Cu(I) in the presence of excess TGA and BA, resulting in the formation of HOOCCH$_{2}$S$\cdot$SCH$_{2}$COOH (di-TGA) as a byproduct. Sn(II) can be oxidized to Sn(IV) by the di-TGA when both Cu(II) and Sn(II) are present in the system, thus creating a self-inhibited redox circle that prevents the valence changes of Cu(I) and Sn(IV) \cite{Guo2020coordination}.
Figure \ref{Liquid Raman} shows Raman spectra of Sn-MCC. Sn-MCC is found to be a mixture of [Sn$_{2}$S$_{6}$]$^{4-}$ and [Sn$_{2}$S$_{7}$]$^{2-}$. The structure of [Sn$_{2}$S$_{6}$]$^{4-}$ exhibits a bridging mode (Sn-S-Sn) and stretching modes at 344 and 367 cm$^{-1}$ respectively, as well as an Sn-S ring vibration mode at 280 cm$^{-1}$. [Sn$_{2}$S$_{7}$]$^{2-}$ was generated by excess sulfur bridges in [Sn$_{2}$S$_{6}$]$^{4-}$ showing characteristic peaks around 300 and 351 cm$^{-1}$ and a new vibrational stretching mode around 348 cm$^{-1}$ 
 \cite{Yang2013SnMCC}. Furthermore, inorganic Sn-MCC complexes were stabilized by the protonated amine from equations \ref{equation 3-2} and \ref{equation 3-3} via electrostatic absorption in an aqueous solution (equation \ref{equation 3-8}. The CZTS nano-inks exhibit the main peak at 336 cm$^{-1}$, as observed in Figure 2(b), which corresponds to the primary vibration mode of CZTS \cite{Fernandes2011Raman}. Several weaker peaks corresponding to Sn-MCC in the solution can still be recognized. No distinct secondary phase was observed, suggesting a pure single-phase constituent in the liquid sample. 

\subsection{High-Quality CZTSSe Absorber Films Made from CZTS Nano Inks}

CZTS precursor films were deposited by spin-coating the nano-inks onto molybdenum-coated glass substrates followed by a soft-baking process for 2 min at 320 $^{o}$C. To avoid the influence of air, all of the above fabrication steps were carried out in an argon-filled glove box. The baking temperature was selected based on thermogravimetric analysis (TGA) and differential thermal analysis (DTA) data that were first carried out on the dried CZTS nano inks, shown in Figure \ref{TGA&XPS&XRD}. The weight loss occurred in the range from 100 $^{o}$C to 320 $^{o}$C. Thus, the as-deposited CZTS precursor thin film was annealed on a preheated hotplate at 320 $^{o}$C.

\begin{figure}[ht]
    \centering
    \includegraphics[width=\textwidth]{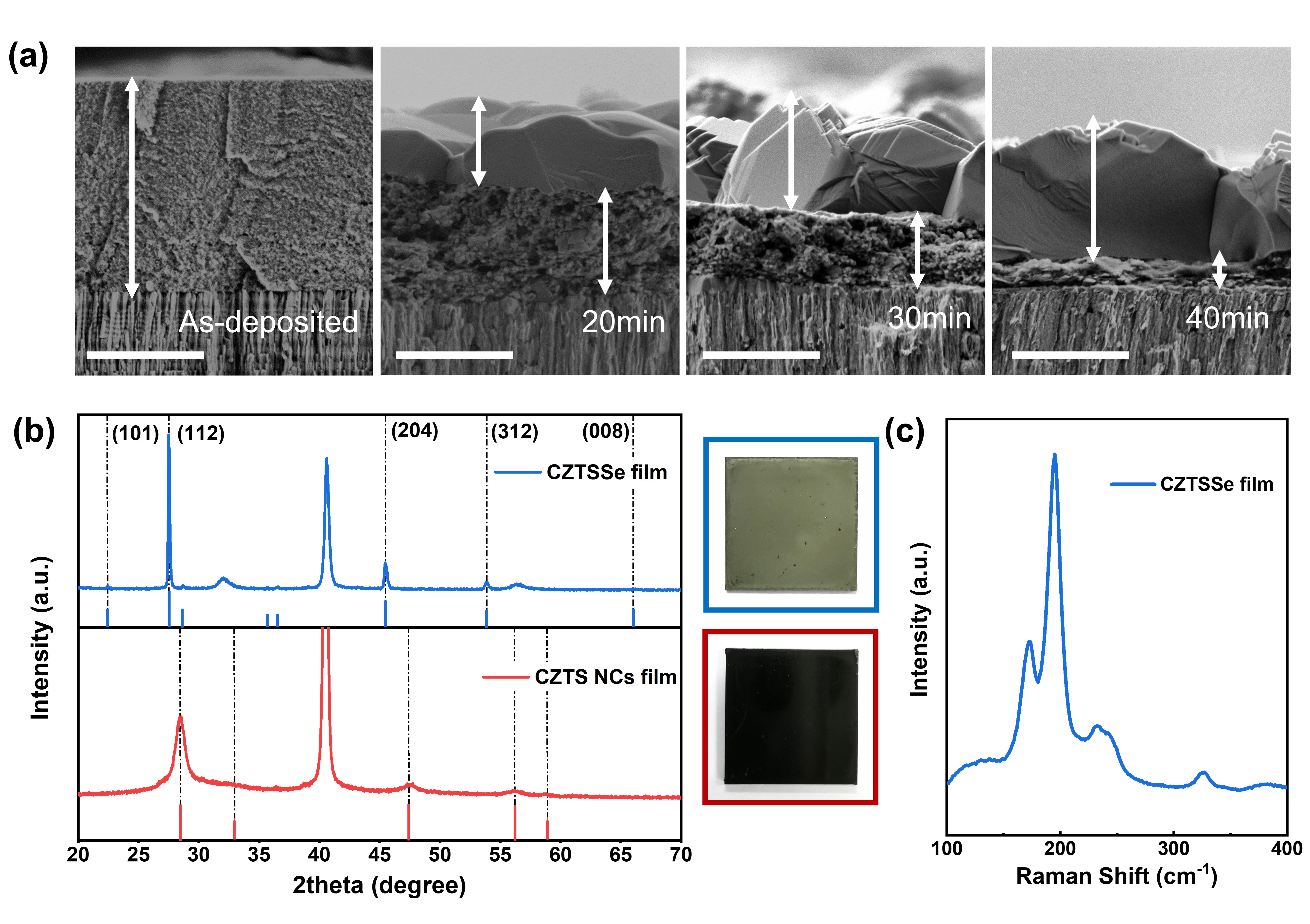}
    \caption[Cross-sectional morphology of CZTSSe absorbers and phase identification.]{(a) SEM images of CZTS films after various selenization durations at 550 $^o$C. Scale bars are 1 $\upmu$m. (b) XRD patterns of an as-deposited CZTS precursor film and a typical selenized CZTSSe film (40 min @ 550 $^o$C) on Mo substrates. Photos of typical CZTS precursor film before (red) and after selenization (blue) are shown for reference. (c) Raman spectrum of selenized CZTSSe film.}
    \label{SEM&XRD&Raman}
\end{figure}

Cross-sectional scanning electron microscope (SEM) images of the morphological changes of the CZTSSe layer under optimized selenization durations are shown in Figure \ref{SEM&XRD&Raman}. The CZTS precursor layers were heated to 550 $^{o}$C and then maintained at this temperature for different periods. A bi-layer structure, consisting of a large grain top layer and a fine grain underlayer, was observed. By extending the duration of the selenization from 20 min to 40 min, the thickness of the small-grain layer was reduced while the large-grain layer grew in thickness. Figure 3.6(b) shows X-ray diffraction patterns of CZTS precursor films soft-baked at 320 $^{o}$C, with no obvious impurity peaks present. The broad (112), (220), and (312) diffraction peaks indicate the formation of small CZTS grains. The selenization process improves the crystallization of the CZTSSe films, resulting in sharper diffraction peaks. The shifting of the diffraction peaks is attributed to the expansion of the lattice as Se becomes incorporated. All of the peaks in the CZTS precursor film and the selenized sample can be assigned to kesterite CZTS (PDF\# 26-0575) and CZTSSe (PDF\#52-0868) respectively. Raman spectroscopy was also performed on selenized films to identify the presence of secondary phases. As shown in Figure 3.6(c), the selenized CZTSSe film showed pure kesterite CZTSe peaks at 172 cm$^{-1}$, 196 cm$^{-1}$, and 238 cm$^{-1}$ and a CZTS peak at 335 cm$^{-1}$ \cite{Joel2014cu2znsns4}. Both XRD and Raman analyses confirm that the CZTSSe films made from synthesized CZTS nano inks do not contain any other secondary phases.

In CZTSSe solar cell research, different ratios of Se and S are widely employed to adjust the band gap. The bandgap of CZTS is around 1.5 eV, which can be gradually tuned to 1.0 eV by replacing S with Se. Figure \ref{Sulfo&Sele} illustrates the changes in the Raman and XRD spectra of the absorber layer under varying sulfurization or selenization conditions. It is evident that with increasing Se content, the main Raman peak of CZTS shifts from 335 cm$^{-1}$ to 196 cm$^{-1}$ for 100$\%$ CZTSe. This shift, induced by an increase in the amount of Se powder used during selenization, indicates the feasibility of adjusting the film composition via the selenization atmosphere. The trends observed in the XRD spectrum also align with the predictions of the Debye-Scherrer equation. The introduction of Se with a larger atomic radius induces expansion of the kesterite unit cell, causing the (112) peak to shift to lower diffraction angles.

\subsection{Phase Evolution Pathway Selected by Using CZTS Molecular Solution and Hybrid CZTS Nano-ink}

\begin{figure}[H]
    \centering
    \includegraphics[width=\textwidth]{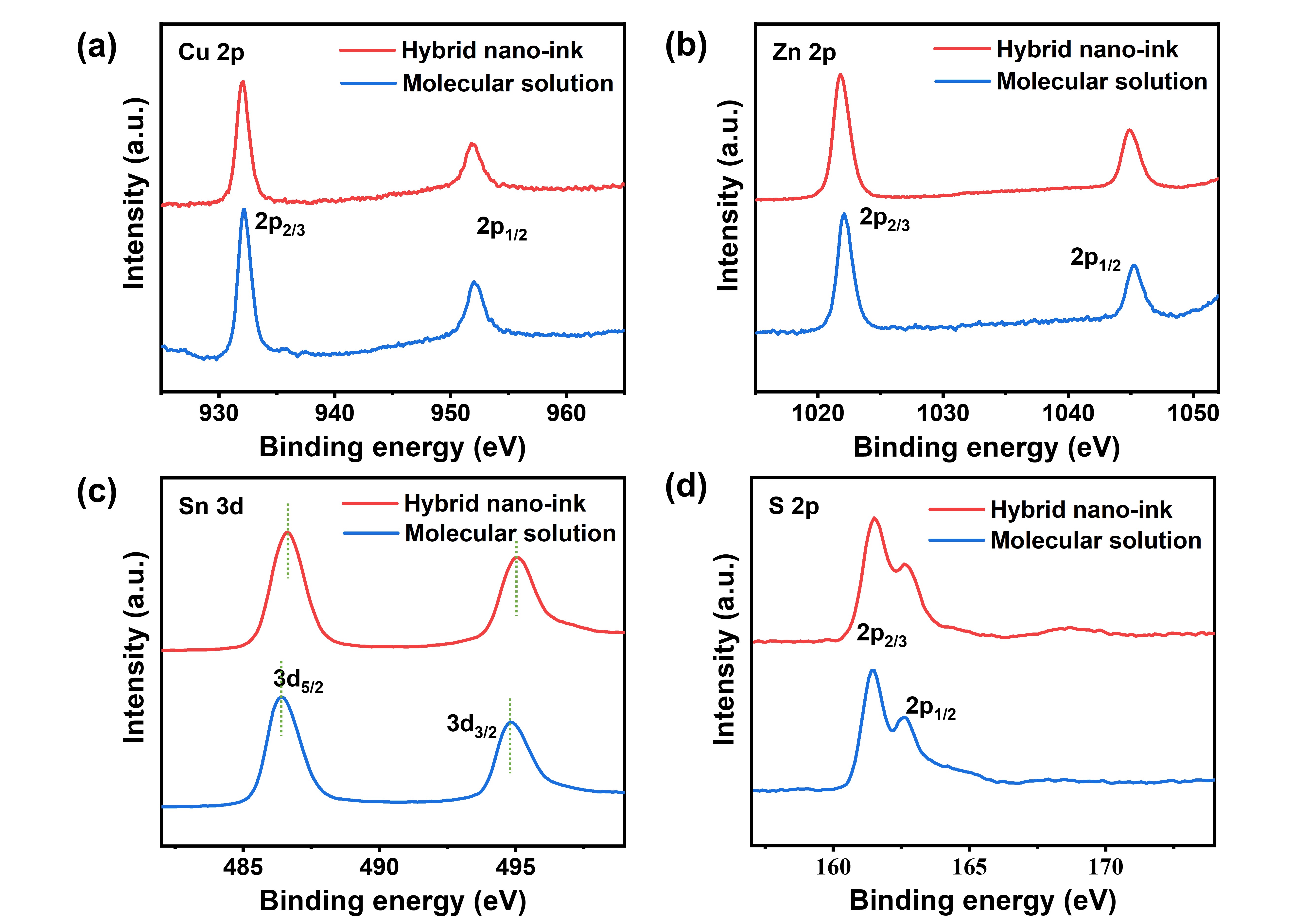}
    \caption[XPS of CZTS films made from molecular solution and hybrid nano inks.]{XPS of precursor films deposited with molecular solution (blue line) and hybrid nano-inks (red line) without high-temperature selenization.}
    \label{XPS}
\end{figure}

Valence states of all cation precursors were investigated by X-ray photoemission spectroscopy (XPS) of spin-coated precursor films before the selenization reaction. A full-range XPS spectrum of the precursor CZTS films deposited using the molecular solution and hybrid nano-ink is shown in Figure \ref{TGA&XPS&XRD}, indicating the presence of Cu, Zn, Sn, and S. High-resolution core level spectra are recorded in Figure \ref{XPS} for the Cu 2p region, Zn 2p region, Sn 3d region, and the S 2p region to determine the valence state. 

Two narrow and symmetric peaks at 933 and 952 eV in the Cu 2p region are assigned to Cu$^{+}$, with no other peaks found at other binding energies, indicating that only Cu$^{+}$ exists in both samples. The typical Zn 2p core level spectrum shows peaks at 1022 and 1045 eV, corresponding to the Zn 2p3/2 and 2p1/2 levels, suggesting the existence of Zn$^{2+}$. The Sn 3d5/2 and 3d3/2 peaks located at 486.6 and 495 eV indicate the presence of pure Sn($^{4+}$. The S 2p core level spectrum shows two peaks, 2p3/2 and 2p1/2, at 162 and 163 eV, which are consistent with the 160-164 eV range expected for S XPS signals in sulfide phases \cite{XPS2010}. In summary, XPS shows that all four elements in both samples made from molecular precursors and from the hybrid nano ink experience no valence changes during film formation, paving the way for wet-chemical synthesis of pure kesterite phase CZTSSe films.

\begin{figure}[ht]
    \centering
    \includegraphics[width=\textwidth]{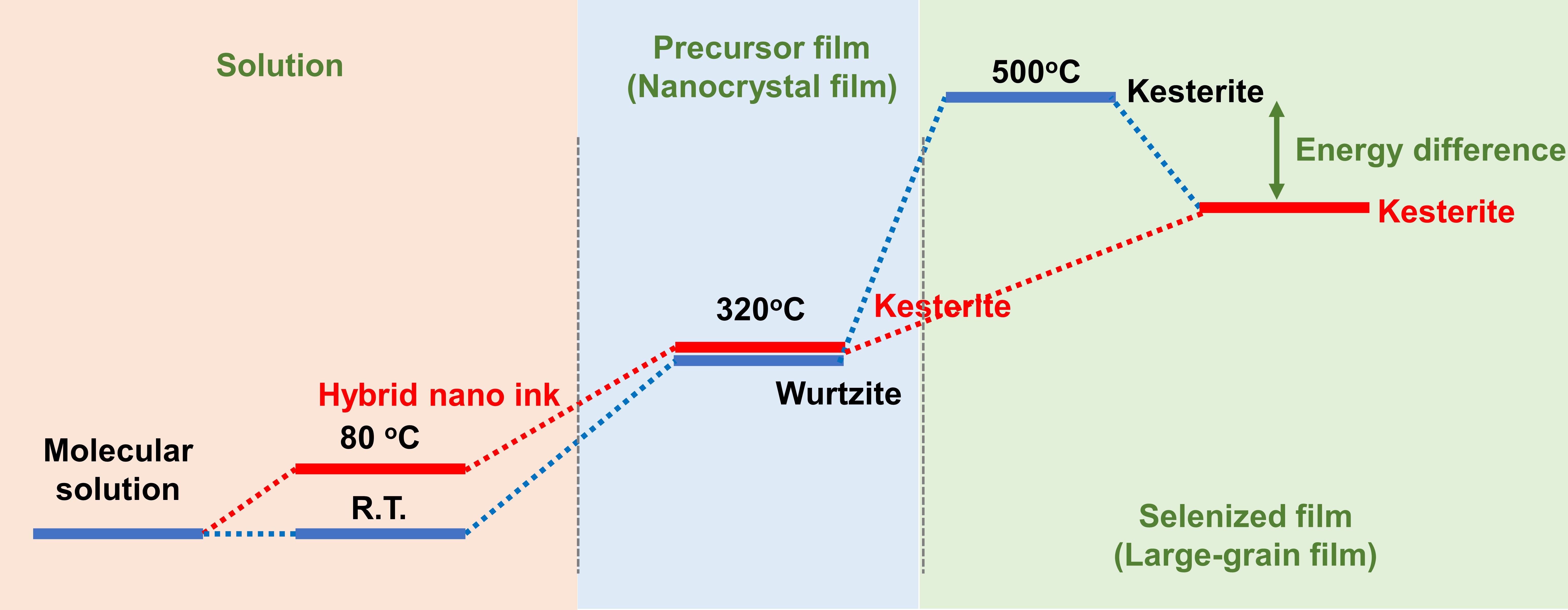}
    \caption[Phase transition and energy roadmap from different synthesis pathways.]{Phase transition and energy roadmap from different synthesis pathways.}
    \label{Phasepath}
\end{figure}

The quantitative presence of metal ions in the correct valence state implies that neither sample contains significant amounts of secondary phases. However, this does not mean that the phase formation pathways are identical. Upon meticulous examination of the transitions from the solution state to the CZTS precursor films, and finally to the selenized CZTSSe films, it becomes evident that the two solutions undergo distinctly different phase transformations, shown in Figure \ref{Phasepath}. In the molecular solution, no significant secondary phase formation is observed throughout the transitions between the three states. However, the wurtzite phase of CZTS is identified in the precursor films. This wurtzite phase is only transformed into the more useful kesterite phase during the subsequent high-temperature selenization process \cite{2014kesteritefromwurtzite}.

\begin{figure}[ht]
    \centering
    \includegraphics[width=\textwidth]{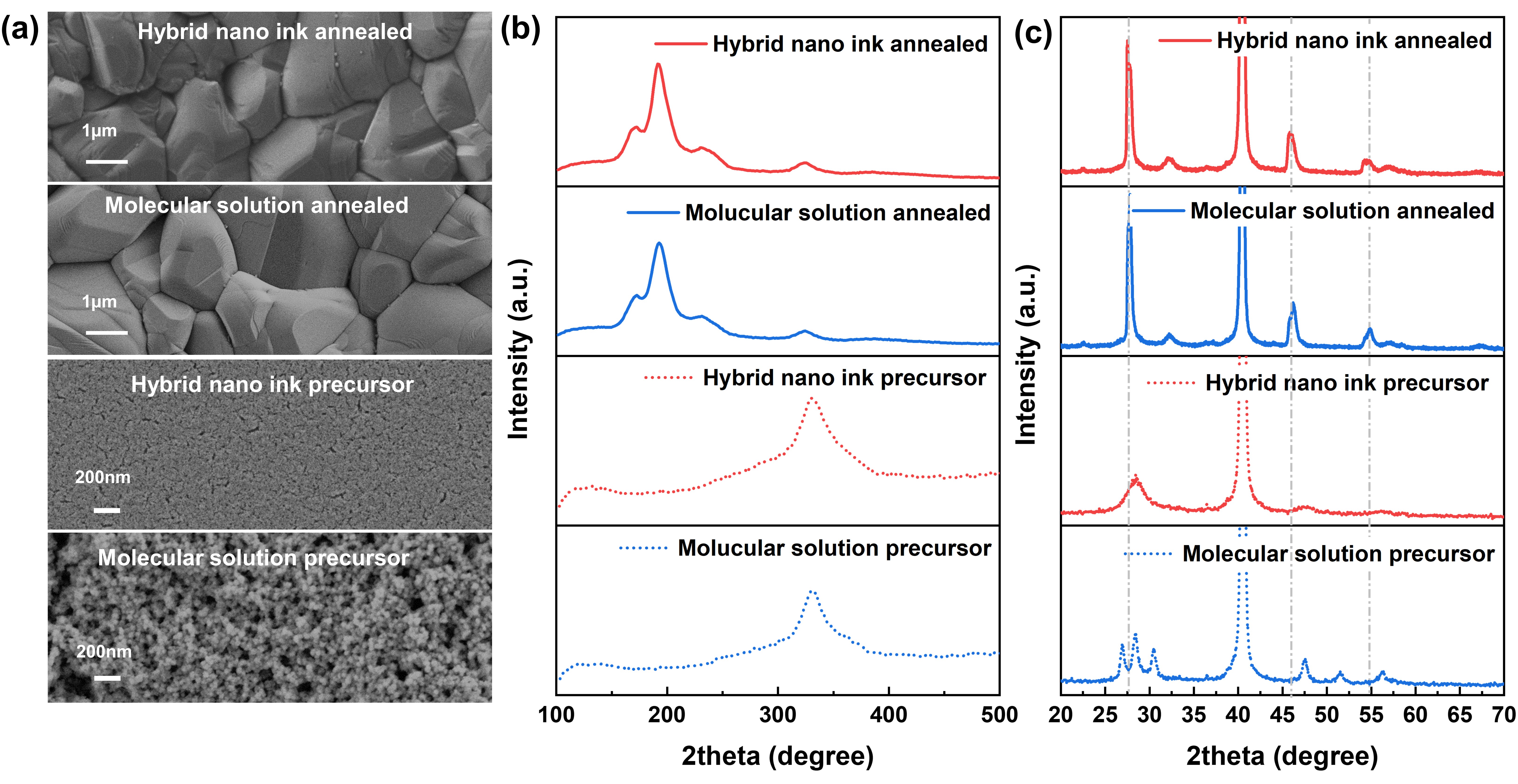}
    \caption[Plane view SEM images, XRD, and Raman of the CZTS and CZTSSe films made from two precursors.]{(a) Plane view SEM images of the precursor (soft baking at 320 $^{o}$C) and selenized CZTSSe films made from molecular solutions and from hybrid nano inks. The corresponding Raman and XRD spectra are displayed in (b) and (c).}
    \label{Phase}
\end{figure}

Normally control over the wurtzite and kesterite phases in CZTS is typically exerted by the choice of elemental precursors used, the order of their addition, and the energy required for synthesis \cite{2020phasecontrolfromthiol,2021phasecontrolfromtemplate, Zou2013phasecontrolfromZnandS, pinto2017controllingsulfur}. The unique aspect of this study lies in the fact that we observe the formation of different CZTS phases simply by altering the reaction temperature over quite a low range (using the same base solution). Detailed analysis of this phenomenon is still pending. However, the key new result is quite critical: the only variable is the reaction temperature, which is around 80 $^o$C, significantly lower than the 200 $^o$C generally required for synthesizing metal sulfide nanoparticles.  Despite this, formation of CZTS nanocrystals can occur directly in solution and when this is the case, conversion appears to be straight to the kesterite phase. Conversely, when the same base solution is employed but the reactants are still molecularly dispersed, nucleation leads to the wurtzite phase. This must be converted into kesterite at a much higher temperature. Hence there are enormous synthetic, energetic, and kinetic advantages to making CZTS films from nao-inks rather than molecular solutions.

The SEM images, Raman, and XRD spectra of CZTS precursor films and selenized CZTSSe films made from two solutions are displayed in Figure \ref{Phase}. A more porous CZTS precursor film is generated from the molecular solution, originating from the decomposition of the dense TGA/Amine solution. Conversely, due to the presence of CZTS nanoparticles in the hybrid nano-ink, the surface stress of the precursor film created via repeated layer-by-layer deposition is released to a great extent, resulting in smoother CZTS precursor films with smaller cracks \cite{Wang2014device}. The distinct phase transition from wurtzite to kesterite can be observed in the sample obtained from a molecular solution using XRD. Furthermore, no other by-products, such as binaries or ternaries, can be easily distinguished at any state from any samples, which helps to exclude the influence of other impurities.

\subsection{Photovoltaic Performances}

\begin{figure}[ht]
    \centering
    \includegraphics[width=\textwidth]{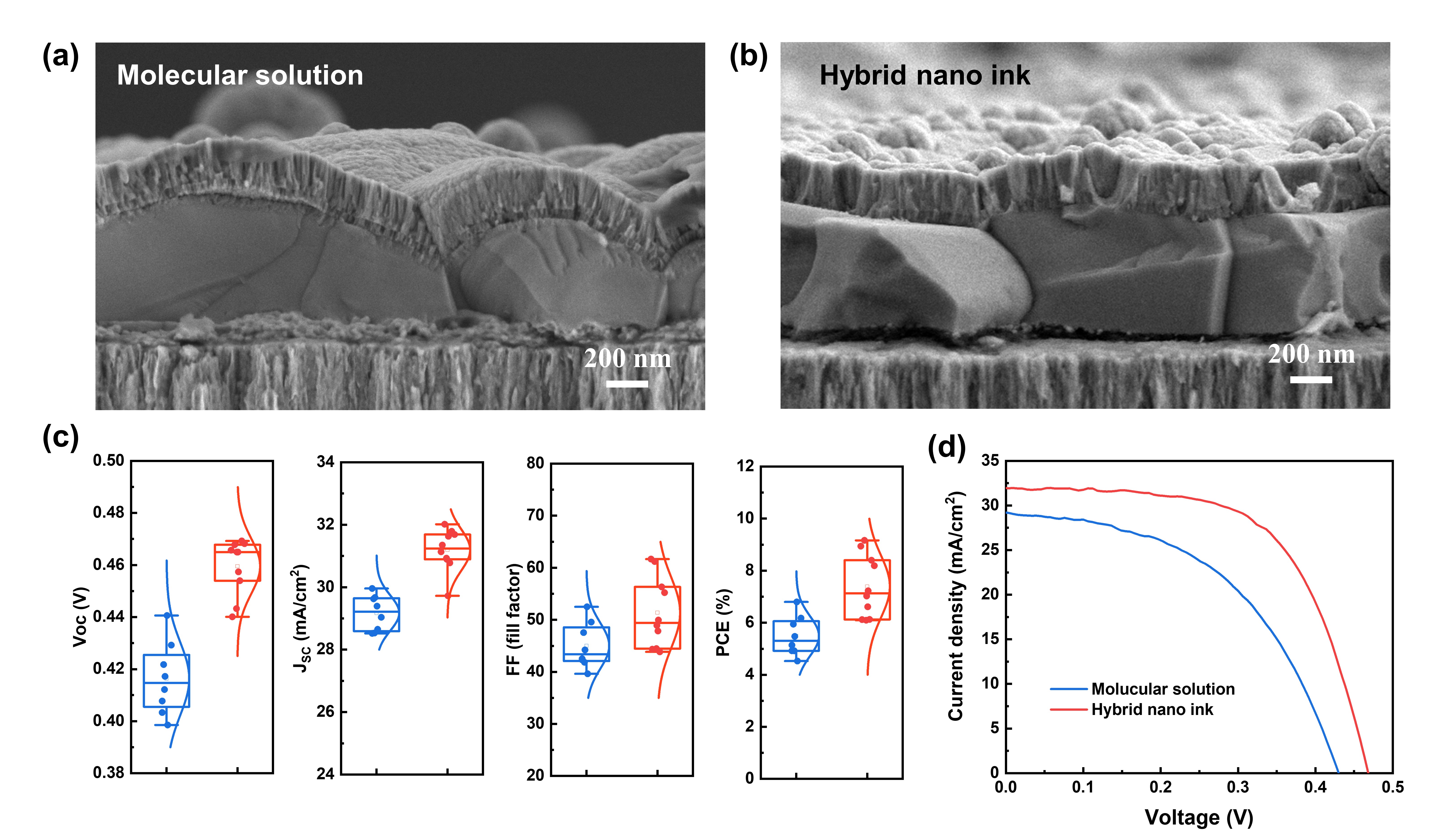}
    \caption[Morphology of CZTSSe solar cells and their PV performance.]{Cross-section SEM image of CZTSSe solar cell fabricated from (a) molecular solution and (b) hybrid nano ink. (c)  Plots of the V$_{oc}$, J$_{sc}$, FF, and power conversion efficiency for 8 independent solar cells. (d) J-V curves for the champion devices.
 XRD patterns. 
}
    \label{PV}
\end{figure}

The photovoltaic performance of as-fabricated CZTSSe solar cells, built from molecular solutions and hybrid nano inks, is presented here. CZTSSe solar cells were fabricated with a device configuration of glass/Mo/CZTSSe/CdS/i-ZnO/ITO/Ag without MgF$_2$ antireflection coating. Glass/Mo substrates were commercially purchased, and CdS was produced by chemical bath deposition. i-ZnO and ITO were deposited by magnetron sputtering. All the active layers are demonstrated by cross-sectional EDS mapping, shown in Figure \ref{EDS}. A comparison of cross-section SEM is provided in Figure \ref{PV}. The hybrid nano ink results in a compact, big-grain layer on top and a thin fine-grain layer at the bottom, while the sample from the molecular solution has a thicker fine-grain layer. The presence of the fine grain is typically attributed to carbon contaminants from residual organic solvents that have not fully decomposed during the thermal treatment of the absorber layer. The strong bond between the long-carbon-chain ligand and the metal atom in the molecular solution makes its removal more difficult than that of the hybrid nano ink, which had undergone ligand exchange to Sn-MCC beforehand.

The photovoltaic performance parameters of CZTSSe devices, including open circuit voltage (V$_{OC}$), short-circuit current density (J$_{SC}$), fill factor (FF), and power conversion efficiency (PCE) are presented here. Devices made by using hybrid nanoink exhibit significant improvement in V$_{OC}$ over the one made form molecular solution, current density, and fill factor, potentially stemming from the pure kesterite phase of the CZTSSe absorber and a direct phase transfer without a meta-stable wutzite phase of CZTSSe. Although the characterization of the solar cells here is limited and further detailed discussion is needed, we can still conclude that avoiding complex phase transition during the synthesis process effectively ensures the quality of the nano ink and the high performance of the solar cells. This provides a new foundation for understanding and optimizing the synthesis process of kesterite and further pushing its efficiency.

\subsection{Conclusion}
\hspace{1em}In conclusion, we have used Cu(I), Zn(II), and Sn(IV) complexes as metal sources to prepare water-based CZTS precursor solutions by redox-free synthesis route. The coordination processes between Cu$^+$/Zn$^{2+}$ and deprotonated SH in TGA/BA, in collaboration with the tin metal chalcogenide complex, enable the facile synthesis of stable and pure kesterite phase nanocrystals in solution. The hybrid nano-ink can be processed by an in situ low-temperature ligand exchange process. Although both the molecular solution and hybrid nano-inks can be used to fabricate CZTS precursor films without side redox reactions, the solid particles in the hybrid nano-inks lead to more compact and defect-free CZTS precursor films. After selenization, the precursor film converts quantitatively into CZTSSe. Assembly of these CZTSSe films into solar cells yielded a full area IPCE efficiency of over 9\%. Notably, cells made by hybrid nano inks demonstrate significantly better performance than those made with molecular solutions, underscoring the importance of the phase evolution pathway in keserite. 

Our approach, which leads to the direct formation of the pure kesterite CZTS phase from metal sources further exemplifies the potential for high-performance, environment-friendly CZTSSe solar cells.

\section{Experimental Procedures}

\subsection{Materials}
\hspace{1em}Copper(I) oxide (Cu$_{2}$O, 99 \%), zinc oxide (ZnO, 99.99 \%), tin powder (Sn, 99.99 \%), sulfur powder (S, 99.99 \%), thioglycolic acid (aqueous solution), butylamine (HO(CH$_{2}$)$_{2}$NH$_{2}$, 99.0 \%), cadmium sulfate (CdSO$_{4}$, ACS), thiourea ((NH$_{2}$)$_{2}$CS, ACS), ammonia solution (NH$_{3}$H$_{2}$O, 25 wt.\%), and ammonium sulfide solution ((NH$_{4}$)$_{2}$S, 40${-}$48 wt.\% in H2O) were purchased from Aladdin. All chemicals and solvents were used as received without any further purification.

\subsection{Synthesis of the CZTS Precursor Solution} 
\hspace{1em}First, 0.45 mmol of Cu$_{2}$O was dissolved in the mixture of 1.5 mL of ethanol, 1.5ml n-butylamine, and 0.5 mL thioglycolic acid aqueous solution in a glass vial under magnetic stirring at room temperature. Next, 0.62 mmol of ZnO was added. In addition, 0.5 mmol of Sn powder, 1 mmol of S powder, and 1 mL of ammonium sulfide solution were mixed in another glass container to achieve a clear solution \cite{Ritchie2018aqueous}. In this work, target ratios of Zn/Sn = 1.24 and Cu/(Zn + Sn) = 0.8 were designed for the CZTS nano-inks. Then the above-mentioned two solutions were mixed to yield a transparent, uniform molecular solution. This was immediately heated up to 80 degrees and stirred for 30 min to get a dark and stable solution. The as-made CZTS ink was then ready for film deposition. No more washing was needed and all the preparation procedures were conducted in air.

\subsection{Deposition of CZTSSe Absorbers}
\hspace{1em}The CZTS nano-inks were transferred into a nitrogen-filled glovebox, and several drops of the solution were spin-cast onto the surface of molybdenum-coated glass at 3000 rpm for 30 s, followed by curing of the film first at 320 $^o$C on a hot plate for 2 min. Such processes were performed several times until the desired thickness was achieved. Typically, 8 layers of solution were coated onto the substrate(~1.6 $\upmu$m). Afterward, the as-prepared CZTS thin film was selenized in a graphite box containing 200 mg Se powder at 550 $^o$C for 10 min or 30 min to form the CZTSSe poly-crystalline film. 

\subsection{Fabrication of CZTSSe Solar-Cell Device}
\hspace{1em}CZTSSe devices with the structure glass/Mo/CZTSSe/CdS/i-ZnO/ITO-/Ag were fabricated. Conventional CdS buffer layers were deposited by the chemical bath deposition (CBD) method. Typically, 5 ml of CdSO$_{4}$ aqueous solution with a concentration of 0.03 mol/L (CdSO$_{4}$, Sigma-Aldrich ACS reagent, 99.0\%), 10 ml of thiourea aqueous solution with a concentration of 0.15 mol/L ((NH$_{2}$)$_{2}$CS, Sigma-Aldrich ACS reagent, 99.0\%), 25 ml of ammonium hydroxide (NH$_{4}$OH, 25\%) and 250 ml deionized water were mixed in a glass reactor. The deposition occurred at 75 $^{o}$C for 15 min to obtain a desired thickness of around 60 nm. I-ZnO (70 nm) and ITO (400 nm) thin films were successively deposited by RF-sputtering. The Ag grid electrode (~300 nm) was made through thermal evaporation. No anti-reflection coating was utilized. Finally, the CZTSSe device with a full area of 0.25 cm$^{2}$ was separated by mechanical scribing.

\subsection{Characterization}
\hspace{1em}Top-view and cross-sectional images were taken using JEOL JSM-7900F field emission scanning electron microscopy (FESEM). The acceleration voltage was 3 kV. The elemental composition was measured with energy dispersive X-ray spectroscopy (EDS) at an electron acceleration voltage of 10 kV on an EDAX Octane X-ray energy spectrometer equipped on JSM-7900F. To verify structural properties and phase purity, X-ray diffraction (XRD) and Raman measurements were conducted by PANalytical Empyrean 2 and inVia Renishaw (532 nm excitation wavelength), respectively. The J-V characteristics of CZTSSe devices were measured using a Xenon-lamp-based solar simulator (Newport, AM 1.5G, 100 mW cm$^{2}$) accompanied by a computerized Keithley 2400 SourceMeter. The illumination intensity of the solar simulator was determined by using a monocrystalline silicon solar cell (Oriel 91150V, $2 cm \times 2 cm$). The external quantum efficiencies (EQE) spectra were measured using a QE-R measurement system (Enli Tech) calibrated by the Enli Tech. Optoelectronic Calibration Lab-certified reference Si and Ge photodiodes.

\begin{acknowledgement}

The authors want to thank Dr.Eser Akinoglu for all the support during the COVID-19. Yixiong Ji thanks the Australian Government and the University of Melbourne for a research training program (RTP) scholarship. This work was funded by the ARC Centre of Excellence in Exciton Science through ARC Grant CE170100026. 

\end{acknowledgement}

\newpage
\begin{suppinfo}

\renewcommand\thefigure{S\arabic{figure}}
\setcounter{figure}{0}

\renewcommand\theequation{S\arabic{equation}}
\setcounter{equation}{0}

Supporting Information is available from the authors.

\begin{figure}[ht]
    \centering
    \includegraphics[width=\textwidth]{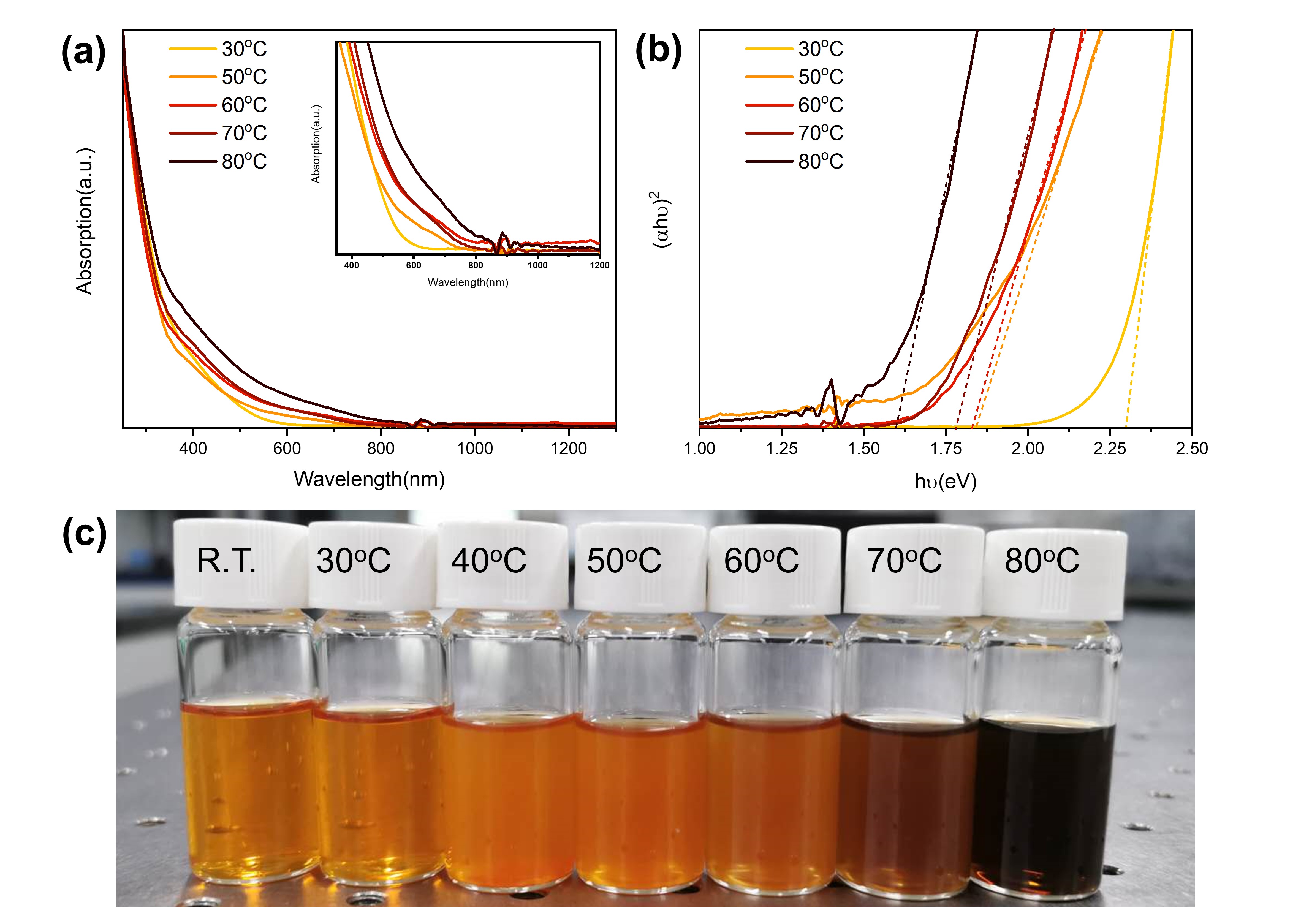}
    \caption[Absorption spectra of the CZTS nano-ink at different temperatures.]{(a) Absorption spectra of the CZTS nano-ink at different temperatures (inset is a partial enlargement of the curve); (b) the corresponding plots of ($\alpha$h$\upsilon$)$^{2}$ vs h$\upsilon$.}
    \label{abs photos}
\end{figure}

\begin{figure}[ht]
    \centering
    \includegraphics[width=\textwidth]{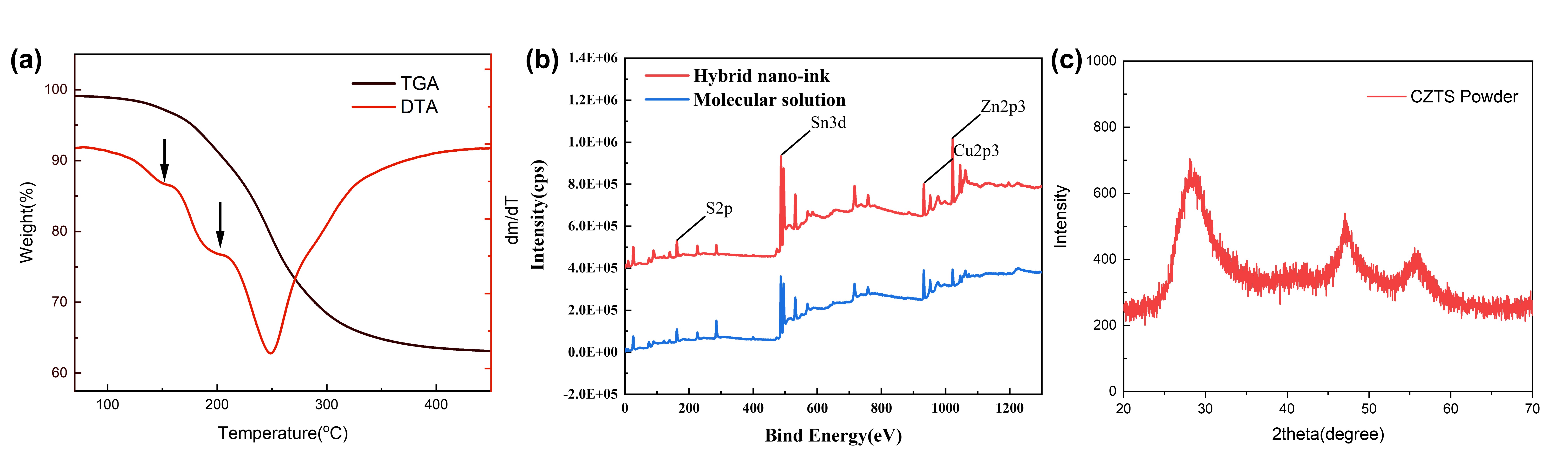}
    \caption[Characterization of as-made CZTS nanoparticle.]{(a) Thermal gravimetric and differential thermal analysis curves of dried CZTS nano inks with a test heating rate of 10 $^{o}$C/min under nitrogen flow.  (b) Full range XPS of hybrid nano-ink (red line) and molecular solution (blue line) (c) XRD of CZTS powder collected from hybrid nano-ink, dried in the glove box at 100 $^{o}$C.}
    \label{TGA&XPS&XRD}
\end{figure}

\begin{figure}[H]
    \centering
    \includegraphics[width=\textwidth]{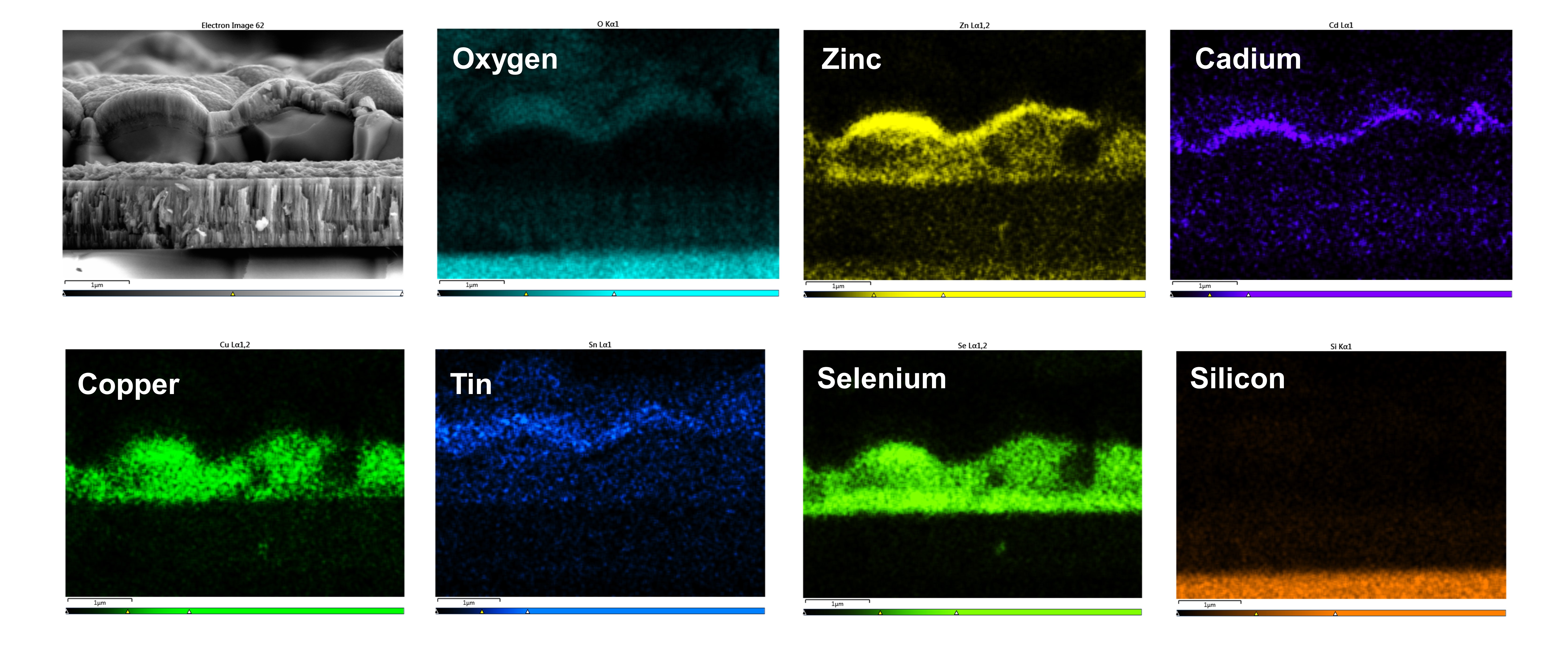}
    \caption[EDS mapping of cross-section CZTSSe solar cells made by using molecular solution.]{EDS mapping of cross-section CZTSSe solar cells made by using molecular solution.}
    \label{EDS}
\end{figure}

\begin{figure}[ht]
    \centering
    \includegraphics[width=\textwidth]{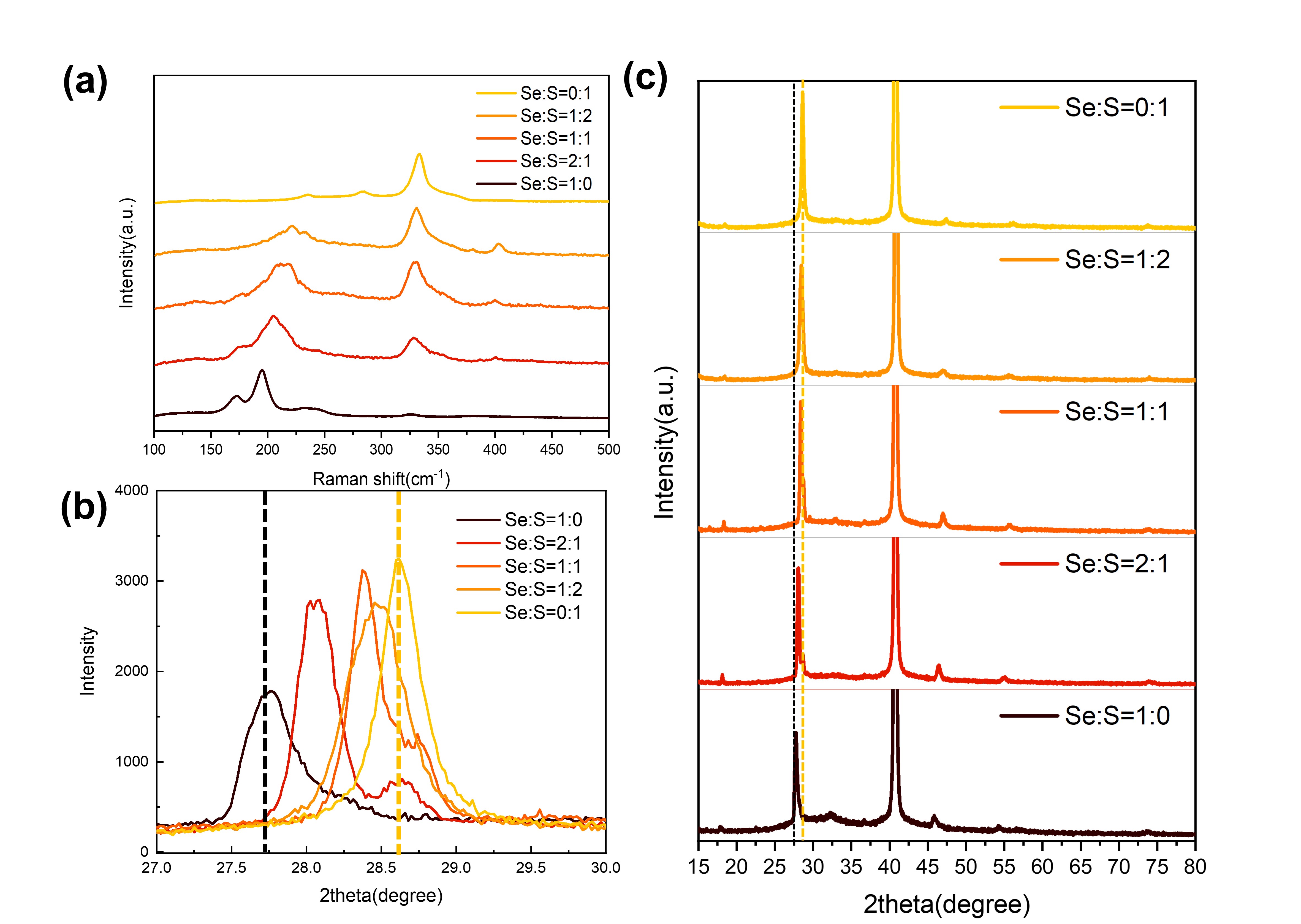}
    \caption[Raman and XRD of Sulfidation and Selenization of CZTSSe films with different ratios of the number of substances between Sulfur and Selenium.]{(a) Raman and (c) XRD of Sulfidation and Selenization of CZTSSe films with different ratios of the amount of substances between Sulfur and Selenium powder (put into the graphite box ). (b) The extracted (112) peaks of CZTS and CZTSSe from XRD patterns.}
    \label{Sulfo&Sele}
\end{figure}

\newpage
\begin{figure}[H]
    \centering
    \includegraphics[width=\textwidth]{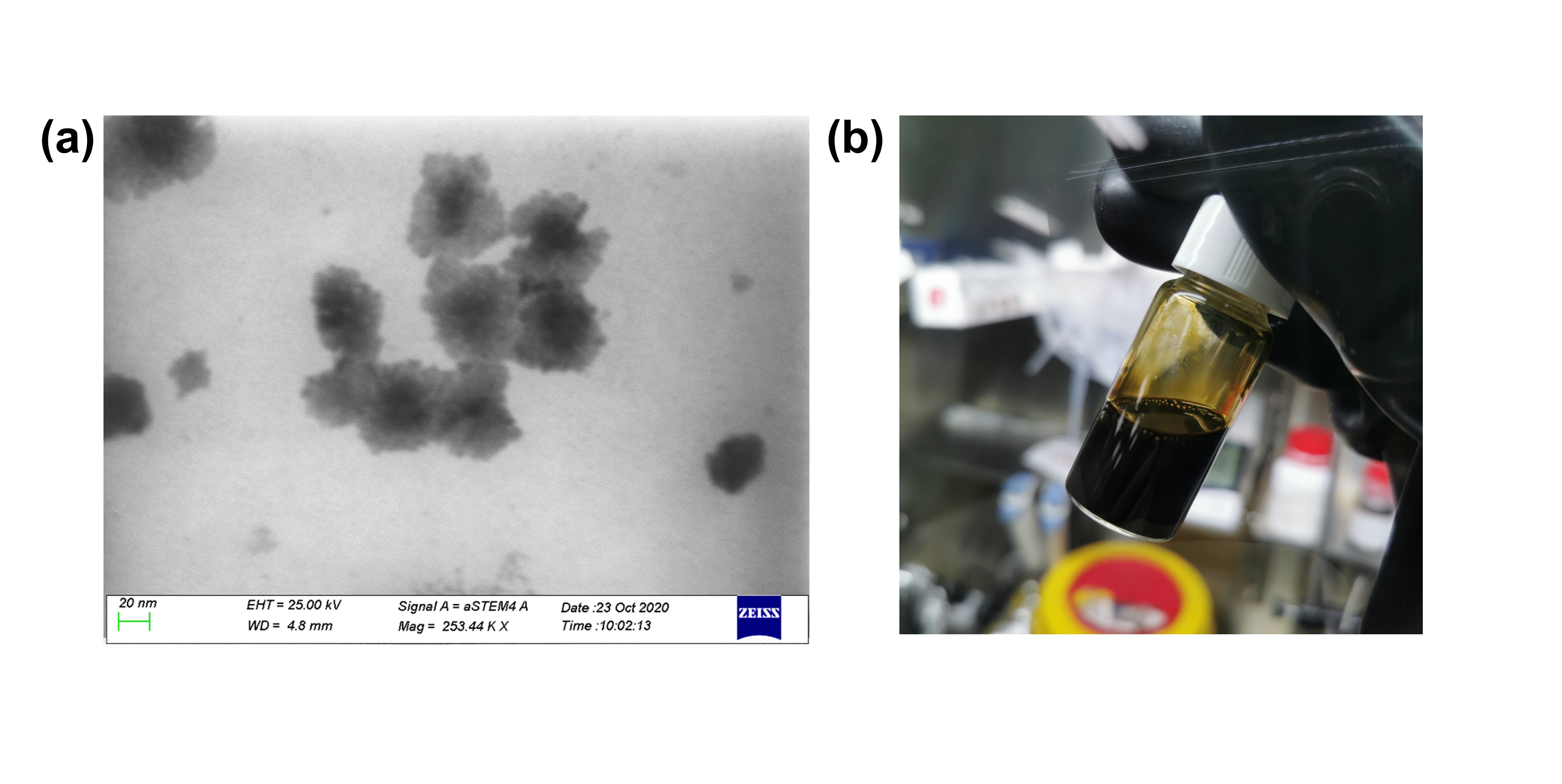}
    \caption[STEM image and camera image of CZTS nanoparticles.]{(a) STEM of the synthesis of CZTS nanoparticle.(b) Photo of synthesized CZTS hybrid nano ink.}
    \label{STEM}
\end{figure}

\end{suppinfo}


\newpage
\bibliography{achemso-demo}

\end{document}